\documentclass[12pt,preprint]{aastex}

\shorttitle{Gravitational Instability of Shear Flows}
\shortauthors{Howes et al.}

\input epsf

\begin{document}

\title{Local Gravitational Instability of Magnetized Shear Flows}

\author{Gregory G. Howes, Steven C. Cowley}
\affil{Department of Physics and Astronomy, University of California,
    Los Angeles, CA 90095-1547}
\email{ghowes@physics.ucla.edu}

\and

\author{James C. McWilliams}
\affil{Department of Atmospheric Sciences, University of California,
    Los Angeles, CA 90095}

\begin{abstract}
The effect of magnetic shear and shear flow on local gravitationally
induced instabilities is investigated.  A simple model is constructed
allowing for an arbitrary entropy gradient and a shear plasma flow in
the Boussinesq approximation.  A transformation to shearing magnetic
coordinates achieves a model with plasma flow along the magnetic field
lines where the coordinate lines are coincident with the field lines.
The solution for the normal modes of the system depends on two
parameters: the Alfv\'en Mach number of the plasma flow and the
entropy gradient.  The behavior of the unstable normal modes of this
system is summarized by a stability diagram.  Important
characteristics of this stability diagram are the following: magnetic
shear is stabilizing and the entropy gradient must exceed a threshold
value for unstable mode growth to occur; flow acts to suppress mode
growth in a substantially unstable regime as expected, yet near
marginal stability it can lessen the stabilizing effect of magnetic
shear and enhance the growth rates of the instability; and, as the
Alfv\'en Mach number approaches one, the instability is completely
stabilized.  Analytical work is presented supporting the
characteristics of the stability diagram and illuminating the physical
mechanisms controlling the behavior of the model. A derivation of the
stability criterion for the case without shear flow, asymptotic
solutions in the limit that the Alfv\'en Mach number approaches one
and in the limit of zero growth rate, a complete WKB solution for
large growth rates, an exactly soluble bounded straight field case,
and energy conservation relations are all presented. The
implications of this work for astrophysical and fusion applications
and the potential for future research extending the results to include
compressibility are discussed.
\end{abstract}


\section{Introduction}

The gravitational stability of a fluid against convective motion has
been extensively studied over the past century.  Pioneering
examinations of the stability of unmagnetized and magnetized
compressible fluids have been conducted by \citet{sch06},
\citet{new61}, and \citet{par79}.  These papers have had a profound
influence on diverse subjects from the dynamics of astrophysical
objects to the confinement of plasma in a fusion device.  Here we
examine the effects of shear in the magnetic field and of an applied
shear plasma flow on stability against gravitational interchange.
These effects change the stability properties and our results are
important for many applications.

We construct a simple model to study the effects of magnetic
shear and shear flow on the stability properties of a magnetized
plasma in a gravitational field.  We derive the equations which
determine the behavior of this model in the Boussinesq limit.  The
equations depend on two parameters, the plasma flow Alfv\'en Mach
number and the entropy gradient.  We conduct a numerical study of the
normal modes of instability and summarize the behavior of the unstable
modes by a stability diagram.  The stability diagram demonstrates
three important characteristics. First, the entropy gradient must
exceed a threshold value for unstable mode growth to occur.
Therefore, the shear magnetic field can stabilize a nonzero entropy
gradient. Second, as expected, shear flow does act to suppress
unstable mode growth when the system is at a substantially unstable
point in parameter space.  But, surprisingly, near marginal stability,
shear flow actually enhances the growth rates of the instability and
also lowers the threshold entropy gradient necessary for instability.
The effect of magnetic shear---to stabilize the plasma and increase
the threshold entropy gradient---is diminished by the addition of
shear flow.  The system can extract energy from the shear flow to
further drive the system to instability.  Third, as the Alfv\'en
Mach number approaches one, the unstable growth rate is suppressed; the
normal modes of the instability are completely stabilized when the
plasma flow exceeds the Alfv\'en speed. Here, the unstable region in
space where a mode can grow moves faster than any perturbation in the
system; any disturbance will be swept downstream out of the unstable
region, leaving behind a stable plasma.

Analytical work helps us to understand the mechanisms responsible for
these characteristics.  First, the stability criterion for the case
without shear flow is derived demonstrating that a shear magnetic
field can support a positive entropy gradient.  Next, asymptotic
solutions demonstrate analytically that stabilization
occurs as the Alfv\'en Mach number approaches one, yet the threshold
entropy gradient for instability goes to zero in the same limit.  In
addition, a complete WKB solution in the limit of a large
growth rate demonstrates both the stabilization by flow at large
growth rates and destabilization near marginal stability.  Then, a
bounded straight field case is solved exactly to show that the
lowering of the threshold entropy gradient with increased shear flow
is a characteristic of plasma flow along the field lines and not
dependent on the magnetic shear in the general model. Finally,
energy conservation relations are derived and analyzed.

In section~\ref{sec:derive}, we describe the model under consideration
and derive the governing system of equations.  The numerical stability
analysis for the general model and the stability diagram are presented
in section~\ref{sec:stab}.  Section~\ref{sec:anal} contains the
analytical results illuminating the characteristics of
the stability diagram.  We describe an exactly soluble, bounded,
straight-field case in section~\ref{sec:str8}.  Finally, in
section~\ref{sec:disc}, the implications of this work on
galactic physics, accretion disk physics, solar physics, and tokamak
confinement are discussed.

\section{Derivation of Equations} \label{sec:derive}
In this section, we derive the equations for linear perturbations of a
vertically stratified atmosphere with shear magnetic field and shear
flow in the high-$\beta$, or Boussinesq, limit. We motivate and apply
a coordinate transformation which casts the problem in its most
natural form.  Investigating the limit of the most unstable modes, we
derive a system of three coupled first-order ordinary differential
equations which capture the lowest order behavior of the model.

\subsection{Setup and Coordinate Transformation} \label{setup}

\begin{figure}
\plotone{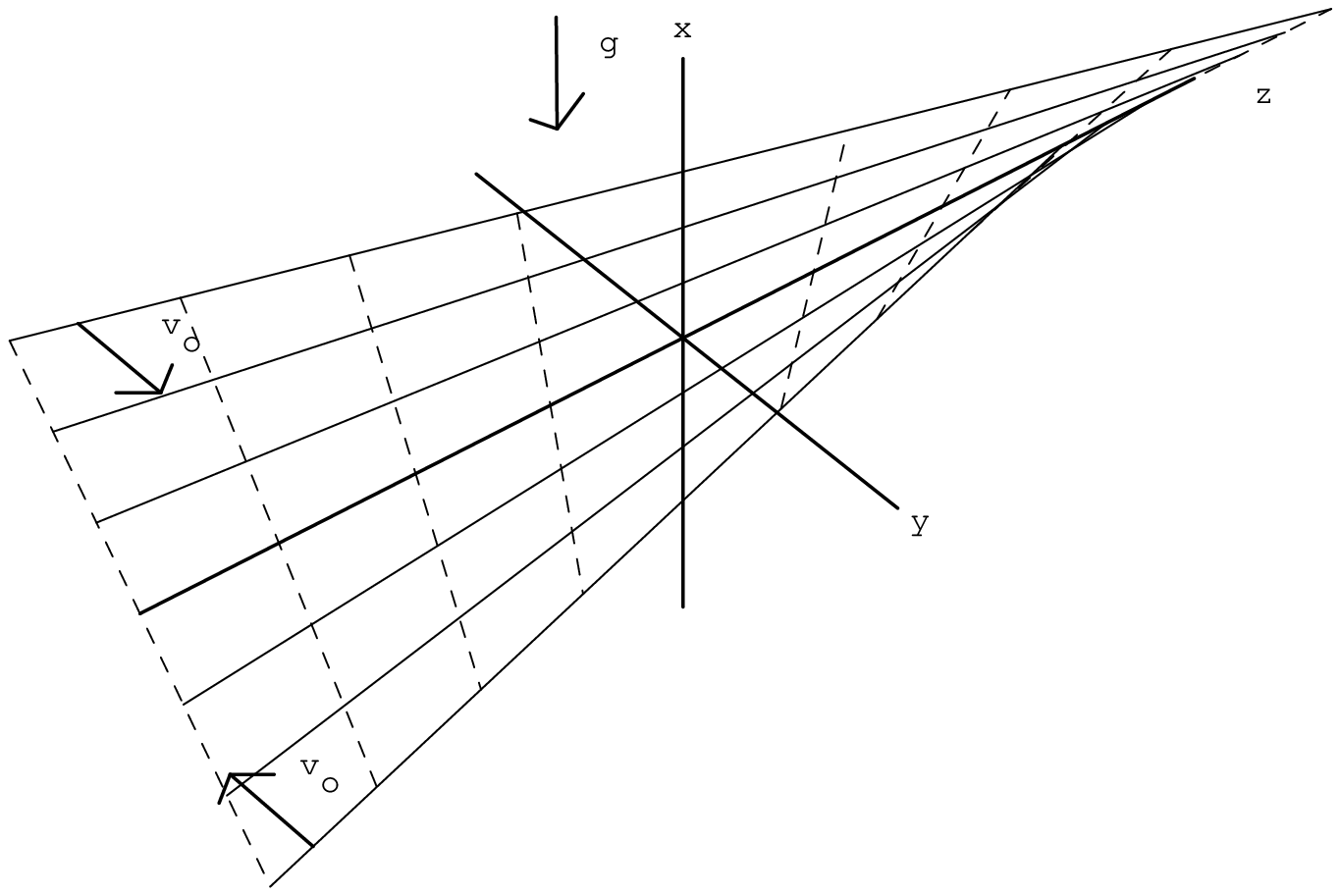}
\caption{The geometry of the shear magnetic field as well as 
the applied shear flow (shown as $v_0$) is shown.  The constant $y'$ surface
is represented by the magnetic field lines (solid lines) and the
dashed lines.}
\label{fig:geom}
\end{figure}

Consider a stationary state for an ideal plasma with mass density
$\rho(x)$ and thermal pressure $p(x)$ and an embedded horizontal
shear magnetic field given by
\begin{equation}
 \mathbf{B}_0 = B_0 \left( \hat{\mathbf{z}} + 
\frac{x}{l_B}\hat{\mathbf{y}} \right) .
\label{eq:B0}
\end{equation}
We then impose a shear flow on this plasma given by
\begin{equation}
 \mathbf{v}_0 = v_0  \frac{x}{l_v}\hat{\mathbf{y}} ,
\label{eq:v0}
\end{equation}
and include gravitational acceleration in the vertical direction given by 
$\mathbf{g} = - g \hat{\mathbf{x}} $. Equilibrium force balance yields
\begin{equation}
 \frac{\partial}{\partial x} \left( p + \frac{B^2}{8 \pi} \right) = - g \rho .
\label{eq:equilib}
\end{equation}
Figure~\ref{fig:geom} shows the geometry of this shear magnetic field
as well as the imposed shear flow on the system.

The instabilities of this plasma are expected to have a short
perpendicular wavelength and a long parallel wavelength (with respect
to the magnetic field) so as to maximize motion in the vertical
direction and minimize field line bending \citep{new61}.  Short
perpendicular wavelengths, however, are rapidly sheared apart by the
perpendicular shear flow.  We would thus like to transform to a coordinate
system with two properties: first, that the flow is along the magnetic
field lines; and, second, that field lines are coordinate lines.
The application of sheared coordinate systems to simplify a problem of
this nature is well documented.  \citet{rob65} employed a coordinate
system in which the field lines are coincident with the coordinate
lines to describe the Rayleigh-Taylor instability of a fluid supported
by a shear magnetic field; \citet{gol65} used a shearing coordinate
system to attack the problem of local gravitational instabilities in a
system with shear flow arising from differential rotation.

We transform the shear velocity to a parallel velocity by transforming to a
frame moving in $z$
\begin{equation}
z'  =  z + v_f t .
\label{eq:ztrans}
\end{equation}
where $v_f= \frac{l_B}{l_v} v_0$.  In this transformed frame, the
velocity becomes
\[
\mathbf{v'}= \mathbf{v_f}+\mathbf{v_0}= v_f \left( \hat{\mathbf{z}} +
 \frac{x}{l_B}\hat{\mathbf{y}} \right) ,
\]
\emph{i.e.} parallel to $\mathbf{B}_0$.
We construct field line coordinates by transforming the $y$ coordinate to
\begin{equation}
 y'  =  y - \frac{x z'}{l_B}  .
\label{eq:ytrans}
\end{equation}
The $x$ coordinate is left unchanged, $x'=x$. It is easy to verify
that $x'$ and $y'$ are constant along field lines ($\mathbf{B_0} \cdot
\nabla x' = \mathbf{B_0} \cdot \nabla y'=0$) and that $\mathbf{B} =
B_0 \nabla x' \times \nabla y'$. The surfaces of constant $y'$ twist
from vertical at $z'=0$ to almost horizontal as $z' \rightarrow \pm
\infty$.  This geometry is shown in Figure~\ref{fig:geom}.

At first, introducing the field line coordinates seems unhelpful since
it introduces explicit $z'$ dependence into the equations.  The
problem also has $x$ dependence that arises from the variation of
$\mathbf{B}$, $p(x)$, and $\rho(x)$.  Indeed, one way to tackle this
problem is to Fourier transform in $y$ and $z$ and solve for the $x$
dependence.  However, the lowest order solution in the twisting
coordinate system is a superposition of these Fourier modes, or,
complementarily, a Fourier solution can be constructed by a superposition
of these twisting modes \citep{rob65}.  We summarize the relationship
between these representations in Appendix~\ref{app:reps}. We consider
the solutions in the twisting coordinates to be more physically
relevant since they are localized in $z'$.

\subsection{Application of Ideal MHD}

The basic equations of ideal MHD include the momentum equation written in
terms of the gradient of total pressure (thermal and magnetic), the
magnetic tension force, and the gravitational force,
\begin{equation}
 \rho \frac{D \mathbf{v}}{D t}
     = - \nabla \left( p+\frac{ \mathbf{B}^2}{8 \pi}  
	\right)
	+ \frac{\mathbf{B} \cdot \nabla \mathbf{B}}{4 \pi} + \rho \mathbf{g},
\label{eq:MHDmom}
\end{equation}
the induction equation in the limit of zero resistivity,
\begin{equation}
 \frac{\partial \mathbf{B}}{\partial t} = \nabla \times (  \mathbf{v} 
   	\times \mathbf{B}) ,
\label{eq:MHDind}
\end{equation}
the continuity equation,
\begin{equation}
 \frac{D \rho}{D t}= - \rho \nabla \cdot \mathbf{v},
\label{eq:MHDcont}
\end{equation}
and the adiabatic equation of state,
\begin{equation}
 \frac{D}{D t} \left( \frac{p}{\rho^\Gamma} \right) =0 .
\label{eq:MHDstate}
\end{equation}
Here $\frac{D}{D t} = \frac{\partial}{\partial t} + \mathbf{v} \cdot \nabla$ 
denotes the Lagrangian derivative, $\mathbf{v}$ 
represents the plasma velocity, and $\Gamma$ is the adiabatic index.
These equations must be evolved subject to the constraint that 
\begin{equation}
 \nabla \cdot \mathbf{B}=0 .
\label{eq:Bdiv}
\end{equation}

Equations~(\ref{eq:MHDmom})-(\ref{eq:MHDstate}), linearized about the
equilibrium (equations~(\ref{eq:B0})-(\ref{eq:equilib})), yield:
\begin{equation}
\left( \gamma' + v_f \frac{\partial}{\partial z'} \right) \delta \mathbf{v}
	+ \delta \mathbf{v} \cdot \nabla \mathbf{v}_0
     = - \frac{1}{\rho_0} \nabla \left( \delta p
	+\frac{ B_0 \delta B_\parallel}{4 \pi}  \right)
    + \frac{B_0}{4 \pi \rho_0} \frac{ \partial \delta \mathbf{B}}{\partial z'}
	+ \frac{\delta \mathbf{B} \cdot \nabla \mathbf{B}_0}{4 \pi \rho_0}
	- \frac{g \delta \rho \hat{\mathbf{x}}}{\rho_0}
\label{eq:MHDmomlin}
\end{equation}
\begin{equation}
\left( \gamma' + v_f \frac{\partial}{\partial z'} \right) \delta \mathbf{B}
	+ \delta \mathbf{v} \cdot \nabla \mathbf{B}_0
     = B_0\frac{ \partial \delta \mathbf{v}}{\partial z'}
	+ \delta \mathbf{B} \cdot \nabla \mathbf{v}_0
	- \mathbf{B}_0 \nabla  \cdot \delta \mathbf{v}
\label{eq:MHDindlin}
\end{equation}
\begin{equation}
\left( \gamma' + v_f \frac{\partial}{\partial z'} \right) \delta \rho
	=  -\delta \mathbf{v} \cdot \nabla \rho_0
	- \rho_0 \nabla  \cdot \delta \mathbf{v}
\label{eq:MHDcontlin}
\end{equation}
\begin{equation}
\left( \gamma' + v_f \frac{\partial}{\partial z'} \right) \delta p
	=  -\delta \mathbf{v} \cdot \nabla p_0
	- \Gamma p_0 \nabla  \cdot \delta \mathbf{v},
\label{eq:MHDstatelin}
\end{equation}
where we have taken all quantities to vary in time as $e^{\gamma't}$.

In a straight field \citep{new61} and a shear field \citep{rob65}
without flow, the most unstable perturbations are incompressible to
lowest order and have a small horizontal perpendicular wavelength.
Such perturbations maximize vertical motion which extracts energy from
the gravitational potential energy and minimize horizontal motions
which extract no energy.  To isolate these motions, we impose the ordering
\begin{mathletters}
\begin{eqnarray}
 \frac{\partial}{\partial y'} = i k &\sim & 
\mathcal{O}\left(\frac{\epsilon^{-1}}{l_B}\right)\\
 \frac{\partial}{\partial x'} & \sim & 
\mathcal{O}\left(\frac{\epsilon^{-1/2}}{l_B}\right)\\ 
 \frac{\partial}{\partial z'} & \sim & 
\mathcal{O}\left(\frac{1}{l_B}\right)\\
 \frac{x'}{l_B}  & \sim & 
\mathcal{O}\left(\epsilon^{1/2}\right).
\label{eq:order}
\end{eqnarray}
\end{mathletters}
where $\epsilon = ( k l_B )^{-1} \ll 1 $ is the ordering parameter or our 
problem. Clearly, all perturbed quantities vary as $e^{i k y'}$. It is also convenient to define the vectors,
\begin{mathletters}
\begin{eqnarray}
 \mathbf{b} & = &\frac{\mathbf{B}_0}{B_0} \\
 \mathbf{e}_\wedge &=& \nabla y' \\
 \mathbf{e}_\perp &  = &\frac{\nabla y' \times\mathbf{B}_0 }{B_0} .
\label{eq:proj}
\end{eqnarray}
\end{mathletters}
The perturbed plasma velocity and magnetic field are projected along
these directions, \emph{i.e.}
\[
\delta \mathbf{v} = \delta v_\perp  \mathbf{e}_\perp 
+  \delta v_\wedge \mathbf{e}_\wedge + \delta v_\parallel \mathbf{b}
\]
\[
\delta \mathbf{B} = \delta B_\perp  \mathbf{e}_\perp 
+  \delta B_\wedge \mathbf{e}_\wedge + \delta B_\parallel \mathbf{b}.
\]
Note that the basis vectors $ \mathbf{e}_\perp$, $\mathbf{e}_\wedge$,
and $\mathbf{b}$ are neither unit vectors nor constant in
space---\emph{e.g.}, $\mathbf{B}_0 \cdot \nabla \mathbf{e}_\wedge = -
\frac{B_0} {l_B}\hat{\mathbf{x}}$.  We expand all perturbed quantities
in powers of $\epsilon^{1/2}$ and denote order as a superscript---for
example, $\delta v_\perp = \sum^\infty_{n=0} \delta v_\perp^{(n)}
\epsilon^{n/2}$.  The ordered, perturbed quantities and operators are
substituted into
equations~(\ref{eq:MHDmomlin})--(\ref{eq:MHDstatelin}).
Equations~(\ref{eq:MHDindlin})--(\ref{eq:MHDstatelin}) at
$\mathcal{O}(\epsilon^{-1})$ and the $\mathbf{e}_\wedge$ projection of
equation~(\ref{eq:MHDindlin}) yield
\begin{equation}
\delta v_\wedge^{(0)} =\delta B_\wedge^{(0)} =0.
\label{eq:wedge}
\end{equation}
Thus, the dominant motion is along the constant $y'$ surfaces in the 
 $\mathbf{e}_\perp $ direction.  
Equations~(\ref{eq:MHDindlin})--(\ref{eq:MHDstatelin}) 
at $\mathcal{O}(\epsilon^{-1/2})$ produce
\begin{equation}
\nabla \cdot \left( \delta v_\perp^{(0)} \mathbf{e}_\perp 
+ \delta v_\wedge^{(1)} \mathbf{e}_\wedge \right) =0.
\label{eq:comp}
\end{equation}
Thus, the perpendicular motion is incompressible to lowest order. 
At  $\mathcal{O}(\epsilon^{-1})$, the $\mathbf{e}_\wedge$ component of 
equation~(\ref{eq:MHDmomlin}) gives
\begin{equation}
\delta p^{(0)} +\frac { B_0 \delta B_\parallel^{(0)}}{4 \pi} =0.
\label{eq:pbal}
\end{equation}
Equation~(\ref{eq:pbal}) expresses the fact that, on the time scales of
interest, pressure balance is achieved across the convective eddies
(in the $\mathbf{e}_\wedge$ direction).  In a subsidiary expansion, we
take the high-$\beta$, or Boussinesq, limit ($\beta = \frac{4 \pi
p_0}{B^2}$).  Thus, equation~(\ref{eq:pbal}) reduces to $\delta p^{(0)}
=0$, and we find from equation~(\ref{eq:MHDstatelin}) that
\begin{equation}
\delta \mathbf{v}^{(0)} \cdot \nabla p_0 = - \Gamma p_0 
( \nabla \cdot \delta \mathbf{v})^{(0)}.
\label{eq:div}
\end{equation}
At $\mathcal{O}(\epsilon^{-1/2})$, the $\mathbf{e}_\wedge$ component
of equation~(\ref{eq:MHDmomlin}) yields $\delta p^{(1)} + \frac { B_0
\delta B_\parallel^{(0)}}{4 \pi} =0$, or taking the high-$\beta$
limit, $\delta p^{(1)} =0$.  The final stability equations are
obtained from the sum and difference of the $\mathbf{e}_\perp$
projections of equations~(\ref{eq:MHDmomlin}) and (\ref{eq:MHDindlin})
at $\mathcal{O}(1)$ and from equation~(\ref{eq:MHDcontlin}) using
equation~(\ref{eq:div}) to substitute for $\nabla \cdot \delta
\mathbf{v}$.  After some tedious but straightforward algebra, we
obtain
\begin{equation}
 - ( 1 - M ) \frac{d A_{+}}{d z} = - \gamma A_{+} + ( 1 + M ) 
	\frac{z}{1 + z^2} A_{-} - \frac {s}{(1 + z^2)^{1/2}} 
\label{eq:A+b}
\end{equation}
\begin{equation}
  ( 1 + M ) \frac{d A_{-}}{d z} = - \gamma A_{-} - ( 1 - M ) 
	\frac{z}{1 + z^2} A_{+} +  \frac {s}{(1 + z^2)^{1/2}} 
\label{eq:A-b}
\end{equation}
\begin{equation}
 M \frac{d s}{d z} =  - \gamma s - \frac{1}{2}  
	\frac{A_{+} - A_{-} }{(1 + z^2)^{1/2}} s_0',
\label{eq:sb}
\end{equation} 
where $A_+$ and $A_-$ are the Els\"asser variables defined by
\begin{equation}
 A_+ = \frac{1}{(1+z^2)^{1/2}} \left(\frac{\delta B_\perp}{B_0} + 
\frac{\delta v_\perp}{v_A} \right) 
\label{eq:A+def}
\end{equation}
\begin{equation}
 A_- = \frac{1}{(1+z^2)^{1/2}} \left(\frac{\delta B_\perp}{B_0} - 
\frac{\delta v_\perp}{v_A} \right) 
\label{eq:A-def}
\end{equation}
and the entropy is given by
\begin{equation}
 s =\frac{g l_B}{v_A^2} \left(\frac{\delta \rho}{\rho_0}\right) .
\label{eq:sdef}
\end{equation}
We have normalized so that $z=z'/l_B$ and $\gamma = \gamma' l_B/v_A$.
Equations~(\ref{eq:A+b})--(\ref{eq:sb}) contain two parameters: the Mach number of the plasma flow with respect to the
Alfv\'en speed ($v_A = \frac{ B_0}{(4 \pi \rho)^{1/2}}$),
\begin{equation}
M=\frac{v_f}{v_A},
\label{eq:Mdef}
\end{equation}
 and the entropy gradient,
\begin{equation}
 s_0' =\frac{g l_B^2}{v_A^2} \left( \frac{\rho_0'}{\rho_0} - \frac{p_0'}
	{ \Gamma p_0}\right) ,
\label{eq:s0'def}
\end{equation}
where the primes denote differentiation by $x$.  With the boundary
conditions that $A_+ \rightarrow 0$, $A_- \rightarrow 0$, and $s
\rightarrow 0$ as $\mid z \mid \rightarrow \pm \infty$,
equations~(\ref{eq:A+b})--(\ref{eq:sb}) define an
eigenvalue problem for $\gamma$.  Solution of the stability equations
yields $\gamma(M,s_0')$.  

Equations~(\ref{eq:A+b})--(\ref{eq:sb}) have a simple physical
interpretation.  $A_+$, the Alfv\'en wave going in the negative $z$
(upstream) direction, travels at the (normalized) speed $1-M$.  The
$A_+$ wave is coupled to the $A_-$ wave by magnetic and velocity shear
(the $A_-$ term in equation~[\ref{eq:A+b}]).  The $A_+$ wave is driven
by gravity via the $s$ term in equation~(\ref{eq:A+b}). $A_-$, the
Alfv\'en wave going in the positive $z$ (downstream) direction,
travels at speed $1+M$, is coupled to $A_+$, and is driven by $s$.
The variable $s$ is proportional to the density or entropy
perturbation and it is driven by both Alfv\'en waves, as seen in
equation~(\ref{eq:sb}).

\section{Stability Analysis}
In this section, we discuss the unstable eigenvalues ($\mathbf{Re} \mbox{ } \gamma >
0$) and eigenfunctions of equations~(\ref{eq:A+b})--(\ref{eq:sb}).  We
have not examined the stable part of the spectrum in detail although
numerical results indicate a continuum along the imaginary $\gamma$
axis.  Two properties of equations~(\ref{eq:A+b})--(\ref{eq:sb}) show
that it is sufficient to examine stability in the region $0 \leq M
\leq 1$. First, note that $\gamma( -M,s_0') = \gamma (M,s_0')$ since
we can map equations~(\ref{eq:A+b})--(\ref{eq:sb}) onto themselves by
the changes $M \rightarrow -M$ , $A_+ \rightarrow A_-$, $A_-
\rightarrow A_+$, $z\rightarrow -z$, and $s \rightarrow s$.  Second,
note that the three asymptotic solutions as $|z| \rightarrow \infty$
are:
\begin{eqnarray}
A_+ \sim  e^{\frac{\gamma z}{1-M}} &\mbox{~~~~~~~~~~~~} s,A_- \sim &
\mathcal{O}\left( \frac{1}{z} e^{\frac{\gamma z}{1-M}} \right) 
\label{eq:asymp1} \\
A_- \sim  e^{\frac{-\gamma z}{1+M}} &\mbox{~~~~~~~~~~~~}  s,A_+ \sim &
\mathcal{O}\left( \frac{1}{z} e^{\frac{-\gamma z}{1+M}} \right) 
\label{eq:asymp2} \\
s \sim  e^{\frac{-\gamma z}{M}} &\mbox{~~~~~~~~~~~~}  A_+,A_- \sim &
\mathcal{O}\left( \frac{1}{z} e^{\frac{-\gamma z}{M}} \right) .
\label{eq:asymp3} 
\end{eqnarray}
If $M>1$, there are no acceptable (decaying) asymptotic solutions as
$z \rightarrow -\infty$ for $\mathbf{Re} \mbox{ } \gamma > 0$.  Thus,
$M>1$ has no unstable eigenmodes.  Physically, this is because all
solutions, even the upstream propagating Alfv\'en wave $A_+$, are
swept downstream.

Without flow, MHD stability equations are self-adjoint \citep{ber58}
and $\gamma^2$ is real.  With flow, no such property is known and
$\gamma^2$ can be complex.  However, in all our solution methods,
$\gamma^2$ has been found to be real for this problem, although we
have not been able to prove that this is rigorously true. The discrete
positive real values for $\gamma$ correspond to unstable growing
modes, and the continuum along the imaginary $\gamma$ axis represents
traveling Alfv\'{e}n waves.

In the following subsection, we present the numerical solution of
equations~(\ref{eq:A+b})--(\ref{eq:sb}).  Various analytical limits
that illuminate the numerical solutions are treated in
subsection~\ref{sec:anal}.  An exactly soluble model with a straight
magnetic field that demonstrates qualitatively similar behavior is
presented in subsection~\ref{sec:str8}. Finally, in
subsection~\ref{sec:energy}, energy constraints on the instability are
discussed.

\subsection{Numerical Solutions}
\label{sec:stab}

We looked for normal mode growth in the system defined by
equations~(\ref{eq:A+b})--(\ref{eq:sb}) using three different
numerical methods.  We directly solved for the eigenvalues $\gamma$ of
this system by matrix solution of the corresponding finite difference
equations using the commercial numerical routine package LAPACK.  We
also found the eigenvalues of the equations to high precision using a
1-D shooting code in $z$ driven by an adaptive step-size, fourth order
Runge-Kutta method with fifth order correction (RK45).  Finally, for
equations~(\ref{eq:A+b})--(\ref{eq:sb}) with $\gamma$ replaced by
$\partial / \partial t$, an initial-value code employing Barton's
method \citep{cen84} for second order accuracy in time was written to
determine the fastest growing mode for any choice of parameters.
Results from all three codes were consistent.

A stability diagram of unstable normal-mode growth rates and stable
regions over the parameter space defined by $M$ and $s_0'$ is
presented in Figure~\ref{fig1}.  As we have already noted, no growing
mode exists for $M>1$.  As well, it is obvious that for the
non-positive values of the entropy gradient, $s_0' \le 0$, there can
be no unstable mode growth since the atmosphere is stably or neutrally
stratified; this is demonstrated by the energy arguments presented in
section~\ref{sec:energy}.  Hence, Figure~\ref{fig1} need only cover
the region of $(M,s_0')$ parameter space defined by $0\le M \le 1$ and
$s_0' > 0$ to include all possible unstable mode growth.

Several features of Figure~\ref{fig1} are important to emphasize.
First, for a system without plasma flow (along the line $M=0$), we see
that the entropy gradient must exceed a threshold value, $s_0'>1/4$,
in order to become unstable when the fluid is supported by a shear
magnetic field.  Second, the qualitative effect of increasing the
plasma flow (increasing $M$) on the instability growth rate depends on
both the growth rate and the plasma flow.  Away from marginal
stability ($\gamma \geq 1$), an increase of the plasma
flow---equivalent to moving along a horizontal line to the right on
the stability diagram---decreases the instability growth rate as one
may expect.  Here, the growing perturbation is sheared out
horizontally so that less gravitational potential energy is extracted
by motions along the constant $y'$ surface, reducing the instability
growth rate.  But near marginal stability ($\gamma < 1$), an increase
in the shear flow effects the instability growth rate differently
depending on the relative orders of the growth rate, $\gamma$, and of
one minus the Alfv\'{e}n Mach number, $1-M$.  The diagonal dotted line
in Figure~\ref{fig1} denotes $\gamma=1-M$.  In region (II) of
the diagram, $\gamma > 1-M$ and flow suppresses the instability.  But,
in region (I), where $\gamma < 1- M$, an increase in the flow actually
enhances the instability growth.  This unexpected result can be
explained with some physical insight.  The point where the instability
can grow is localized at $z'=0$ in our transformed coordinates; this
is where the constant $y'$ surface is vertical and motions along that
surface can extract the most gravitational potential energy with which
to drive the instability.  When a shear plasma flow is introduced into
the system, this is manifested in our transformed system by a plasma
flow along the field lines in the $+z'$ direction. This causes the
Alfv\'{e}n modes in the $+z'$ and $-z'$ directions to propagate at
different speeds in our transformed system. The counter-propagating
mode ($A_+$ in our model) is slowed down, spending more time in the
region around $z'=0$ where unstable mode growth occurs.  Hence, the
instability is enhanced by a shear flow in the plasma.  The final
point to be gleaned from Figure~\ref{fig1} concerns the behavior as $M
\rightarrow 1$, or as the Alfv\'{e}n Mach number approaches one. In
this region, every contour corresponding to a finite growth rate
asymptotes to $s_0'\rightarrow \infty$; hence, the instability is
stabilized as the Alfv\'{e}n Mach number approaches one.

\subsection{Analytical Limits}\label{sec:anal}

The stability diagram, Figure~\ref{fig1}, prominently displays the
three main characteristics discussed in section~\ref{sec:stab}: a
threshold entropy gradient, $s_0'>1/4$, necessary for instability in
the system without plasma flow; for increasing plasma flow, a
reduction of the unstable growth away from marginal stability, but an
enhancement of that unstable mode near marginal stability including a
decrease in the threshold entropy gradient necessary for instability;
and the stabilization of unstable normal modes as the Alfv\'{e}n Mach
number approaches one.  Each of these characteristics is relevant
in a different region of the $(M,s_0')$ parameter space of the
diagram.  By examining the model in each of these regions of parameter
space, we can confirm and explain our results analytically.

First, we examine the criterion for instability in the absence of
plasma flow in section~\ref{sec:stabcrit}; this corresponds to the
left vertical axis of the stability diagram where $M=0$.  Next, in
section~\ref{sec:M1lim}, we conduct an asymptotic analysis in the $M
\rightarrow 1$ limit---region (II) of the stability diagram---to
show that the plasma is indeed stabilized as the Alfv\'{e}nic Mach
number approaches one.  In section~\ref{sec:g0}, we investigate the
reduction of the threshold for instability with plasma flow; this
covers the lower, right-hand side of region (I) of the stability diagram.  A
WKB analysis for a large instability growth rate, presented in
section~\ref{sec:wkb}, yields the behavior of the system in the
central and upper portion of the stability diagram; the suppression of
the growth rate by flow in region (II) and its enhancement in region
(I) are verified by the eigenvalue condition $\gamma (M,s_0')$
obtained in this analysis.

\subsubsection{Stability Criterion without Flow}
\label{sec:stabcrit}

Here we obtain the stability criterion for a magnetized fluid
supported by a shear magnetic field in the Boussinesq limit with no
shear flow.  By using the substitution $ v = \delta v_\perp / (1 +
z^2)$, the equations without plasma flow ($M=0$) can be simplified to
a Sturm-Liouville equation of the form
\begin{equation}
 \frac{d}{d z} \left[ (1 + z^2) \frac{d v}{d z} \right] 
	- \left[ \gamma^2 (1 + z^2) - s'_0 \right] v = 0
\label{eq:vm0}
\end{equation}
over the interval $(- \infty, + \infty)$.  The boundary conditions on
this system necessitate that $v \rightarrow 0$ as $z \rightarrow \pm
\infty$.  From Sturm's First Comparison Theorem \citep{inc26}, we
know that, as the eigenvalue $\gamma^2$ is increased, the solution
will oscillate less rapidly with zeros of the function $v$ leaving the
interval $ - \infty < z < \infty$ at the boundaries.  Thus, if the
solution with $\gamma^2 =0$ oscillates, we can increase $\gamma^2$
until the boundary conditions are satisfied, so there will be an
unstable solution.  Note also that the fastest growing mode has no
zeros in the interval and must be even in $z$.

Let w(z) satisfy equation~(\ref{eq:vm0}) with $\gamma^2 =0$.  Substituting
a series solution of the form
\begin{equation}
 w(z) = \sum^{\infty}_{n=0} a_n (1 + z^2)^{-(n+\alpha)}.
\label{eq:vsoln}
\end{equation}
in equation~(\ref{eq:vm0}) (with  $\gamma^2 =0$), we obtain the recurrence
relation,
\begin{equation}
\frac{a_n}{a_{n-1}} = \frac{ 4 ( n + \alpha -1 )^2}
{4 ( n + \alpha -1/4)^2 + s_0' - 1/4},
\label{eq:recur}
\end{equation}
with
\begin{equation}
 \alpha = \frac{1}{4} \pm  \frac{1}{4} \sqrt{ 1-4 s'_0}.
\label{eq:alpha}
\end{equation}

When $s_0' > 1/4$, solution~(\ref{eq:vsoln}) oscillates
for $z \rightarrow \infty$, and clearly an unstable solution exists.
Let us therefore consider stability when $s_0' < 1/4$.  We take the
positive sign in equation~(\ref{eq:alpha}); then $w^2$ is
integrable for $z \rightarrow \infty$.  We note that all $a_n$ are
positive if $a_0$ is positive.  We choose $a_0 > 0$ such that $w>0$.
The series solution for $w(z)$ given by equation~(\ref{eq:vsoln}) is
non-differentiable at $z=0$ (as we see below).  Thus, we cannot use
the solution from equation~(\ref{eq:vsoln}) over the whole interval
and must restrict its use to $z>0$.  Let us suppose (for
contradiction) that there exists at least one unstable solution of
equation~(\ref{eq:vm0}).  Further, let $v_0(z)$ be the most unstable
solution---as noted above, $v_0(z)$ must be even in $z$ and have no
zeros in the interval $ - \infty < z < \infty$.  We therefore choose $v_0(z)>0$
everywhere.  It is straight forward to show that
\begin{equation}
v_0(0)  \left. \frac{d w }{d z} \right|_{0^+} = \gamma_0^2 \int^\infty_{0^+} 
w v_0(1+z^2) dz,
\label{eq:zero}
\end{equation}
where the limit $0^+$ is infinitesimally above $z=0$.  Since the
integral and $v_0$ in equation~(\ref{eq:zero}) are positive, we have
stability, $\gamma_0^2 < 0$ (a contradiction), if $ \left. \frac{d w
}{d z} \right|_{0^+} <0$.  Since every term in the series
equation~(\ref{eq:vsoln}) is a monotonically decreasing function of
$z$ we expect $ \left. \frac{d w }{d z} \right|_{0^+} <0$, but since
the limit is nonuniform we take a more careful approach.  We determine
the sign of the limit $ \left. \frac{d w }{d z} \right|_{0^+}$ from an
examination of the convergence of the series.  It is straightforward
to show that
\begin{equation}
a_n \sim \frac{A}{n^{3/2}}
\label{eq:anlim}
\end{equation}
as $n \rightarrow \infty$ with $A$ a positive constant.  Thus, the
series for $w(z)$ (equation~[\ref{eq:vsoln}]) converges for $ z \geq
0$.  However, the series for $\frac{d w }{d z}$ converges for $z>0$
but not for $z=0$.  Let us write for $z \rightarrow 0$
\begin{equation}
\frac{d w }{d z} = \sum^\infty_{n=0} \frac{ -2 z(n+\alpha) a_n}
{(1+z^2)^{(n+\alpha+1)}} \simeq  - C z - 2 \sum^\infty_{n=N} \frac{A z n^{-1/2}}
{(1+z^2)^n},
\label{eq:derivlim}
\end{equation}
where $C$ is a positive constant and $N$ is a large number in the
range $1 \ll N \ll z^{-2}$.  Using $(1+z^2)^{-n} \simeq e^{-nz^2}$, we
obtain
\begin{eqnarray}
\frac{dw}{dz} & \simeq  - C z - 2 A \sum^{\infty}_{n=N} n^{-1/2} z  e^{-n z^2}
 \simeq  -C z - 2A \int^{\infty}_{N} n^{-1/2} z  e^{-n z^2} dn 
=  - C z - 2 A\int^{\infty}_{\sqrt{N}z}  e^{-p^2} dp \nonumber \\
& \simeq - C z - 2 A \int^{\infty}_{0}  e^{-p^2} dp
 = - C z - A \sqrt{\pi}.
\label{eq:dvdz2}
\end{eqnarray}
Thus, the limit of $\frac{d w }{d z}$ as $z \rightarrow 0^+$ is
$-\sqrt{\pi} A$, \emph{ i.e.} negative.  From equation~(\ref{eq:zero})
we conclude that $\gamma_0^2 < 0$ and there are no unstable modes for
$s_0' < 1/4$.  Thus, the necessary and sufficient condition for
instability is $s_0' > 1/4$.

This criterion can also be written
\begin{equation}
 \frac{\rho_0'}{\rho_0} - \frac{p_0'}{ \Gamma p_0} >  
\frac {1}{4} \frac {v_A^2}{g l_B^2} .
\label{eq:bouscriterion}
\end{equation}
This confirms the result in Figure~\ref{fig1}---that a threshold
value of the entropy gradient, given by
Equation~(\ref{eq:bouscriterion}), must be exceeded in order to cause
instability when the fluid is supported by a shear magnetic field.
Clearly, without magnetic shear ($l_B \rightarrow \infty$), the usual
criterion, $s_0' >0$, holds and the motion is the simple interchange of
field lines.  With magnetic shear, the field lines \emph{must} be bent
since the interchange of field lines is impossible with finite
displacements---thus magnetic shear is stabilizing.

\subsubsection{Asymptotic Solution in the $M \rightarrow 1$ Limit}
\label{sec:M1lim}
An asymptotic, boundary layer analysis can be carried 
out in the limit that $M \rightarrow 1$.  This
asymptotic expansion is described in Appendix~\ref{app:Mto1}.
The eigenvalue condition derived for this limit in that appendix is
\[
\gamma^2 \simeq s_0' (1-M^2).
\]
This relation explains the upturn towards infinity of the constant
growth rate contours in region (II) of Figure~\ref{fig1} as $M
\rightarrow 1$.

\subsubsection{Solution in the $\gamma \rightarrow 0$, $M \rightarrow 1$ Limit}
\label{sec:g0}
We demonstrate that the threshold entropy gradient for instability
(the limit that $\gamma \rightarrow 0$) decreases as the plasma flow
velocity is increased and that the threshold value of $s_0'$
approaches zero linearly as $M \rightarrow 1$.  Letting $1-M \sim
\epsilon$, we can redefine the following variables in terms of
$\epsilon$: $\gamma = \epsilon \overline{\gamma} $, $ A_-= \epsilon
\overline{A}_-$, $s = \epsilon \overline{s}$ and $s_0' = \epsilon
\overline{s_0'} $.  For $M \rightarrow 1$, we can drop terms of order
$\epsilon^2$ and cancel $\epsilon$ from each remaining term to yield
the simplified set of equations
\begin{mathletters}
\begin{eqnarray}
 -  \frac{d A_{+}}{d z} & = &- \overline{\gamma} A_{+} + 2 
	\frac{z}{1 + z^2} \overline{A}_- - 
\frac {\overline{s}}{(1 + z^2)^{1/2}} \\
 2 \frac{d \overline{A}_-}{d z}& =& - \frac{z}{1 + z^2} A_{+} +  
\frac {\overline{s}}{(1 + z^2)^{1/2}} \\
  \frac{d \overline{s}}{d z}& =& - \frac{1}{2}  
	\frac{A_{+} }{(1 + z^2)^{1/2}}\overline{s_0'}.
\end{eqnarray} 
\end{mathletters}
These equations are now independent of $\epsilon$, or, equivalently,
are independent of $M$.  The equations will hold true for constant
values of $\overline{\gamma}$ and $\overline{s_0'}$.  In this case, if
we have a negligibly small value $\overline{\gamma} \rightarrow 0$, we obtain
the corresponding threshold value of $s_0' =\overline{s_0'}(1-M) $, where
$\overline{s_0'}$ is a constant.  Thus, the threshold entropy gradient
for stability must linearly approach zero as $M \rightarrow 1$ , as
seen in the lower right-hand corner of Figure~\ref{fig1}.

\subsubsection{WKB Analysis}
\label{sec:wkb}
A complete solution of the model for a large
growth rate $\gamma$ can be constructed if we assume an 
ordering, for a small parameter $\epsilon$, such 
that $ \gamma \sim \mathcal{O}( \epsilon^{-1})$,  $ \frac{d}{dz} \sim 
\mathcal{O}( \epsilon^{-1})$, $ s \sim \mathcal{O}( \epsilon^{-1})$,
and $ s'_0 \sim \mathcal{O}( \epsilon^{-2})$.  In this case, we can
neglect the second term on the right hand side of both
equations~(\ref{eq:A+b}) and (\ref{eq:A-b}).  Converting back
from Els\"{a}sser variables to $\delta B_\perp$ and $\delta v_\perp$
notation, combining the three equations into a single second order
equation, and neglecting the term $M s \frac{d}{dz}
\left( \frac{1}{(1+z^2)^{1/2}} \right)$
(which can be shown to be small), yields
\begin{equation}
(1-M^2) \frac{d^2 \delta v_\perp}{dz^2} - 2 \gamma M \frac{d \delta
v_\perp}{dz} - \left[ \gamma^2 - \frac{s_0'}{1 + z^2} \right] \delta
v_\perp = 0.
\label{eq:wkb1}
\end{equation}  
Changing variables with an integrating factor to $v = \exp \left(
 \frac{ \gamma M z } {1 - M^2} \right) \delta v_\perp$, we obtain
\begin{equation}
 \frac{d^2 v}{dz^2} -  \frac{1}{(1-M^2)^2}
 \left[ \gamma^2 - \frac{s_0'(1-M^2)}{1 + z^2} \right] v = 0.
\label{eq:wkb2}
\end{equation}  

We assume a WKB solution of the form $e^{i \int k(z) dz}$ and find
\begin{equation}
v= \overline{v} \exp \left({\pm \frac{i}{1-M^2} \int^z \left[
\frac{s_0' (1-M^2)} {1+z'^2} - \gamma^2 \right]^{1/2} dz'} \right).
\label{eq:wkbsols}
\end{equation}  
The turning points are at $z= \pm z_0$ where
\begin{equation}
z_0^2= \frac{s'_0 ( 1-M^2)}{\gamma^2} -1 . 
\label{eq:z0}
\end{equation}  For $\mid z \mid > z_0$, the WKB solutions are decaying
exponentials; in the region $ -z_0 < z<z_0$, the WKB solution is an
oscillatory function.

In the usual way \citep{ben78}, we obtain the Bohr-Sommerfeld
quantization condition
\begin{equation}
\int^{z_0}_{-z_0} \left[ \frac{s'_0 ( 1-M^2)}{1+z^2} - \gamma^2 \right]^{1/2}
	dz = 2 n \pi (1-M^2).
\label{eq:BScond}
\end{equation} 
The growth rate of the $n$th mode is then given, for small $z_0$, by
\begin{equation}
\gamma^2 = \left[ s_0' - 2 n (s_0')^{1/2}(1-M^2)^{1/2} \right]
(1-M^2).
\label{eq:gammawkb}
\end{equation} 
For a large growth rate $\gamma \sim \mathcal{O}( \epsilon^{-1})$ and the lowest, nontrivial eigenmode $n=1$, we can solve this 
eigenvalue condition for $s_0'$ to obtain
\begin{equation}
s_0'= \frac{\gamma^2}{1-M^2} + 2 \gamma.
\label{eq:s0wkb1}
\end{equation} 
This condition agrees with the behavior of the constant growth rate
contours in region (II) of Figure~\ref{fig1}.  Although we do not
expect the eigenvalue condition in the WKB approximation to be precise
in the limit of small growth rate, $\gamma \sim \mathcal{O}( \epsilon) $, we do find to lowest order the qualitatively correct
form,
\begin{equation}
s_0' \sim 1-M^2,
\label{eq:s0wkb2}
\end{equation} 
that the contours in region (I) decrease like $1-M^2$.

\subsection{Bounded, Straight-Field Case}
\label{sec:str8}
The general model defined by this paper has the characteristic that
the shear in the magnetic field localizes the region of instability
around where the constant $y'$ surface is vertical (this corresponds
to $z'=0$ in our transformed coordinate system).  With this
characteristic as our guide, an exactly soluble, simplified model can
be constructed which demonstrates the same qualitative behavior
displayed in Figure~\ref{fig1}.  We construct a case with a straight
magnetic field and a plasma flow along the field lines which has
boundaries at $z=\pm L$.  Since the explicit $z$ dependence drops out
of the equations in the straight-field limit ($l_B \rightarrow
\infty$), we Fourier transform in $z$ to obtain the algebraic
dispersion relation
\begin{equation}
( \gamma + i M k_z ) \left[ ( 1- M^2) k_z^2 + 2 i \gamma M k_z + \gamma^2
-s_0' \right] = 0.
\label{eq:simpdisp}
\end{equation} 
If we remove the plasma flow from the system by setting $M=0$,
this dispersion relation agrees with the results of \citet{new61} for a
straight-field without flow in the Boussinesq limit.  

We solve for the three solutions for $k_z$ from
equation~(\ref{eq:simpdisp}) and find the eigenvectors corresponding
to each $k_z$.  Constructing general solutions for $\delta B_\perp$,
$\delta v_\perp$, and $s$ from these eigenvectors, we find the
eigenvalue condition $\gamma(M,s_0')$ that must hold in order to
satisfy the three necessary boundary conditions on the system.  The
boundary conditions we apply are $\delta v_\perp=0$ at $z = \pm L$
and the upstream boundary condition $s=0$ at $z = - L$.  The
eigenvalue condition thus obtained is
\begin{equation}
 \gamma ^2 = \left[ s_0' - \left(\frac{ n \pi}{ 2 L} \right)^2 (1 - M^2) 
\right] (1-M^2)
\label{eq:simpeig}
\end{equation} 
for the $n$th order unstable mode where $n = 1,2, 3, \ldots$.

A plot of constant $\gamma$ contours is displayed in
Figure~\ref{fig:simp}.  Note that the qualitative behavior pointed out
in the text in section~\ref{sec:stab} is demonstrated by this
simplified model.  Therefore, the magnetic shear is not responsible
for the unexpected decrease in the stability threshold with shear
flow; only a localization of the instability is necessary to
demonstrate this characteristic.  A quantitative comparison of
Figures~\ref{fig1} and \ref{fig:simp} shows that, to yield equivalent
growth rates, a much larger entropy gradient must be supplied in the
unbounded case with magnetic shear than in the bounded, straight-field
model.  We can understand the difference as follows.  The energy
required to bend the magnetic field lines slows the growth of the
instability.  In the bounded case, this energy is needed to bend the
field line only within the bounded domain.  But, in the unbounded
model, the bending of the field lines occurs over a larger extent in
$z$, thus requiring more energy and so more effectively suppressing
the instability.

\subsection{Energy Conservation}
\label{sec:energy}
In the standard way, equations~(\ref{eq:A+b})--(\ref{eq:sb}) can be
combined, replacing $\gamma$ with the time derivative
$\frac{\partial}{\partial t}$, to obtain an energy integral for the
model.  Converting back to $\delta v_\perp$ and $\delta B_\perp$ 
using equations~(\ref{eq:A+def}) and (\ref{eq:A-def}), we find
\begin{equation}
\frac{\partial}{\partial t} \left\{ \frac{1}{2} \int^\infty_{- \infty} \left[
	\frac{ \delta B_\perp^2 + \delta v_\perp^2}{1 + z^2} - \frac{2 M \delta B_\perp \delta v_\perp}
	{1 + z^2} - \frac{s^2}{s_0'} \right] dz \right\} = M 
	\int^\infty_{- \infty} \frac{ s \delta B_\perp} {1 + z^2} dz.
\label{eq:Bousenergy}
\end{equation} 
Adopting the terminology of \citet{hay87}, we define the integral on
the left-hand side of equation~(\ref{eq:Bousenergy}) as the wave
energy of the perturbation.  Note that, in the absence of flow
($M=0$), the wave energy is constant in time.
Equation~(\ref{eq:Bousenergy}) supplies a limit on the value of $s_0'$
necessary for instability: since the $\delta B_\perp^2$ and $\delta
v_\perp^2$ terms are both positive definite, an instability can only
develop for $s_0'>0$.  In this case, gravitational potential energy
from the $s^2$ term can be harnessed to drive the kinetic energy and
field line bending of the instability.

By the same method as above, we find that the wave energy integral for
the straight-field model bounded at $z= \pm L$, in
section~\ref{sec:str8}, has the form
\begin{equation}
\frac{\partial}{\partial t} \left\{ \frac{1}{2} \int^L_{-L} \left[
\delta B_\perp^2 + \delta v_\perp^2 - \frac{s^2}{s_0'} \right] dz \right\} =\frac{M}{2}
\left\{  \delta B_\perp^2(-L) - \delta B_\perp^2(L) + \frac{s^2(L)}{s_0'} \right\} .
\label{eq:Straightenergy}
\end{equation} 
Without flow, again, we find that the necessary condition for
instability to develop is $s_0'>0$ and that the wave energy integral
is constant in time.  In the presence of flow, we interpret the terms
on the right-hand side of equation~(\ref{eq:Straightenergy}) as follows:
$\delta B_\perp^2(-L)$ represents the flow of magnetic energy into the
region, $- \delta B_\perp^2(L)$ represents the flow of magnetic energy
out of the region, and $\frac{s^2(L)}{s_0'}$ represents the flow of
gravitational potential energy out of the the region.

\section{Discussion} \label{sec:disc}

To study the effect of magnetic shear and shear flow on local
gravitationally induced instabilities, we have constructed a simple
model in the Boussinesq limit of ideal MHD.  Numerical solutions to
this model yield a stability diagram of the $(M,s_0')$
parameter space.  This stability diagram has three important
characteristics. First, there exists a threshold entropy gradient for unstable mode
growth, demonstrating that magnetic shear is a stabilizing influence. Second,
flow serves to suppress mode growth in a substantially
unstable regime, but near marginal stability it lessens the
stabilizing effect of magnetic shear, enhancing unstable mode growth
rates and lowering the threshold entropy gradient necessary for
instability. Third, normal modes of instability are stabilized completely
as the Alfv\'en Mach number approaches one because the disturbance is
swept downstream out of the unstable region.  Analytical work
corroborates these characteristics in the different regions of $(M,s_0')$
parameter space.

In a shear magnetic field without flow, the characteristic shape of
the unstable mode is such that the field lines remain on the constant
$y'$ surface shown in Figure~\ref{fig:geom}; hence, in the lab frame,
the field lines must twist as they fall under gravity to remain on
this surface. This occurs because the perturbed field line at any
vertical height $x$ must align with the direction of the unperturbed
field at that height to facilitate interchange.  Unlike the ordinary
interchange of straight field lines, if magnetic shear is present, the
field line \emph{must} be bent to allow interchange; this is the root
of the stabilizing influence of the shear magnetic field.  Energy
extracted from gravitational potential energy as the field line falls
must supply both the kinetic energy of the moving plasma, which is
frozen to the field line, and the energy required to bend the field
line.  Unstable motions are localized about the point where the
constant $y'$ surface is vertical ($z=0$ in Figure~\ref{fig:geom}),
because this point is where motions on that surface extract the most
gravitational potential energy while minimizing field line bending.
When plasma shear flow is included in the problem, this unstable
region moves in the lab frame with the flow velocity $v_f$ in the
$-\hat{\mathbf{z}}$ direction.  Therefore, this characteristic
twisting geometry propagates along $z$ at a speed
$v_f=v_0\frac{l_B}{l_v}$ determined by the local flow speed $v_0$,
flow shear length $l_v$, and magnetic field shear length $l_B$.
Typical eigenfunctions for an entropy gradient of $s_0'=0.6$ and
Alfv\'en Mach numbers $M=0$ and $M=0.8$ are displayed in
Figure~\ref{fig:eigfunc}.

The modifications of the gravitational stability of a magnetized
plasma due to the presence of magnetic shear and shear flow impact
several diverse subjects.  In solar physics, one of the key processes
necessary for the success of the interface dynamo \citep{par93,cha96}
is the storage of an intense toroidal field in the solar tachocline
\citep{spi92} until an instability causes an isolated flux tube to
rise into the base of the convection zone
\citep{ada78,cal83a,cal83b,hug87,cat88,cat90,sch94,cal95,gil97,bar98}. 
Shear flow is present due to the large differential rotation of the
sun in the tachocline region \citep{tho96,sch98}, and some magnetic
shear probably exists here as well; the impact of shear flow and
magnetic shear may alter gravitational stability within this region.
There are two regions in galactic physics where the conditions in our
model may apply: near the center of the galaxy, where there likely
exists a region of magnetic shear in which the large-scale magnetic
field changes from primarily azimuthal to primarily vertical
\citep{zwe97} and shear flow is present through differential rotation,
and in the disk of the galaxy, where both magnetic shear and shear flow
may be present in the vertical direction away from the mid-plane.  In
examining the mechanisms for turbulence in accretion disks,
\citet{bal91} pointed out that a previously discovered
\citep{vel59,cha60,cha61} but unappreciated linear MHD instability, 
driven by differential rotation coupled with magnetic tension, would
occur if the accretion disk were threaded by a weak magnetic field.
But the Balbus-Hawley instability bends the magnetic field lines
extensively; thus, it is stabilized by magnetic tension for large
magnetic field strengths \citep{bla94,urp96,kit97}.  Although the
twisting interchange instability studied here depends not on rotation
but on an entropy gradient, it has a characteristic geometry that
minimizes field line bending and so may be important in regions of
large field strength.  As is well known within the fusion community
\citep{rob65}, employing a sheared toroidal magnetic field in a
tokamak can help to stabilize, or at least suppress, ballooning
instabilities (buoyant interchange instabilities driven by pressure
and curvature forces).  The idea of employing shear flow to further
stabilize ballooning instabilities has gained much attention in the
past decade
\citep{wae91,has91,has96,has99,mil95}.  Our work demonstrates that,
near marginal stability, shear flow actually lessens the stabilizing
effect of magnetic shear, lowering the threshold entropy gradient
required for instability and enhancing the unstable mode growth rate.
Full stabilization of the plasma will occur only if the Alfv\'en Mach
number of the plasma flow (as defined in our transformed coordinates)
exceeds one.  The local nature of the instability examined in this
work means that our treatment may apply locally in more complicated
environments.  The instability may behave as a traveling ``wave
packet'' which moves with the intersection of shear magnetic field
lines (where the constant $y'$ surface is vertical); the disturbance
may move in and out of unstable regions, with the perturbation growing
where conditions are unstable and decaying in stable regions.

We have extended the model to include compressibility.  In this case,
the behavior is governed by a system of five coupled, first-order
ordinary differential equations.  Five parameters are necessary to
describe the system: the Alfv\'en Mach number, the plasma $\beta$, the
density gradient, the pressure gradient, and the magnetic field
gradient. Over some portions of this five-dimensional parameter space,
the growth rate eigenvalue $\gamma$ does indeed become complex.
Producing a simple answer from this more complicated model is quite
difficult.  Our current research is addressing this difficulty.  But,
the magnetic-buoyancy instability \citep{par79}, which depends on
compressibility, cannot be investigated without employing this more
detailed treatment.  We will then be able to relate our work
quantitatively to applications such as the stability of magnetic
fields in the solar tachocline. One final possible extension of this
research is an investigation of the nonlinear behavior using an
initial-value code.

%

\acknowledgments
The research was performed under appointment of Greg Howes to the
Fusion Energy Sciences Fellowship program administered by the Oak
Ridge Institute for Science and Education under a contract between the
U.S. Department of Energy and the Oak Ridge Associated Universities.
It was supported in part by the National Science Foundation under
Grant No. PHY94-07194.  We are grateful to B. Albright, A. Cumming,
B. Dorland,  E. Quataert, S. Tobias, J. Toomre, and E. Zweibel for
useful discussions.

\appendix

\section{Twisting Modes vs. Fourier Modes}
\label{app:reps}
In the sheared magnetic field, the local unstable modes can be written
in two ways \citep{rob65}; we call them twisting and
Fourier modes.  For the reader's convenience, we summarize the essence
of the argument here.  In the twisted field line coordinates, we
obtain the twisted mode where a perturbed quantity $\Phi(x,y,z,t)$ is
given by
\begin{equation}
\Phi_T = \overline{\Phi}_T(z,x) e^{iky' +\gamma t} = \overline{\Phi}_T(z,x)
e^{iky - ik \frac{x z}{l_B} +\gamma t}
\label{eq:t1}
\end{equation}
with $\overline{\Phi}_T$ localized in $z$ and varying weakly in $x$
compared to $k^{-1}$.  But since the origin in $z$ is arbitrary, we can
also write
\begin{equation}
\Phi_T' = \overline{\Phi}_T(z-z_0,x) 
e^{iky - ik \frac{x(z-z_0)}{l_B} +\gamma t}.
\label{eq:t2}
\end{equation}
Thus, there are an infinite number of twisting modes, each with a
different origin of the twist.  We can construct a mode that does not
depend on $z$---a Fourier mode---by integrating equation~(\ref{eq:t2})
over $z_0$
\begin{equation}
\Phi_F = \int^\infty_{-\infty} \Phi_T '(z-z_0,x,y,t)dz_0 =
\overline{\Phi}_F \left(\frac{kx}{l_B},x\right) e^{iky +\gamma t},
\label{eq:t3}
\end{equation}
where 
\begin{equation}
\overline{\Phi}_F \left(\frac{kx}{l_B},x\right) = \int^\infty_{-\infty}  
\overline{\Phi}_T(z',x) e^{-i\frac{kx}{l_B} z' dz'}.
\label{eq:t4}
\end{equation}
Thus, the Fourier modes and the twisting modes are related by a
Fourier transform.  Clearly, the Fourier mode can be made by
``adding'' twisting modes together (equation~[\ref{eq:t3}]), or vice
versa, using the Fourier inversion theorem on equation~(\ref{eq:t4});
see \citet{cow91} for pictures of this superposition. The Fourier
modes are narrowly localized in $x$---typically $\Delta x \sim
\frac{\Delta x}{k \Delta z}$ where $\Delta x $ is the $x$ width of the
Fourier mode and $\Delta z$ is the $z$ width of the twisting mode.
Note this $x$ localization of the Fourier mode is narrow compared to
the $x$ variation of the twisting mode.  In this paper, we have taken
the twisting mode representation for two reasons: first, the role of
the flow, we believe, is more intuitive in this picture; and, second,
the twisting modes are finite in $z$ extent and therefore represent
more easily the evolution of an initial value problem.

\section{Asymptotic Analysis as $M \rightarrow 1$}
\label{app:Mto1}

\begin{figure}
\plotone{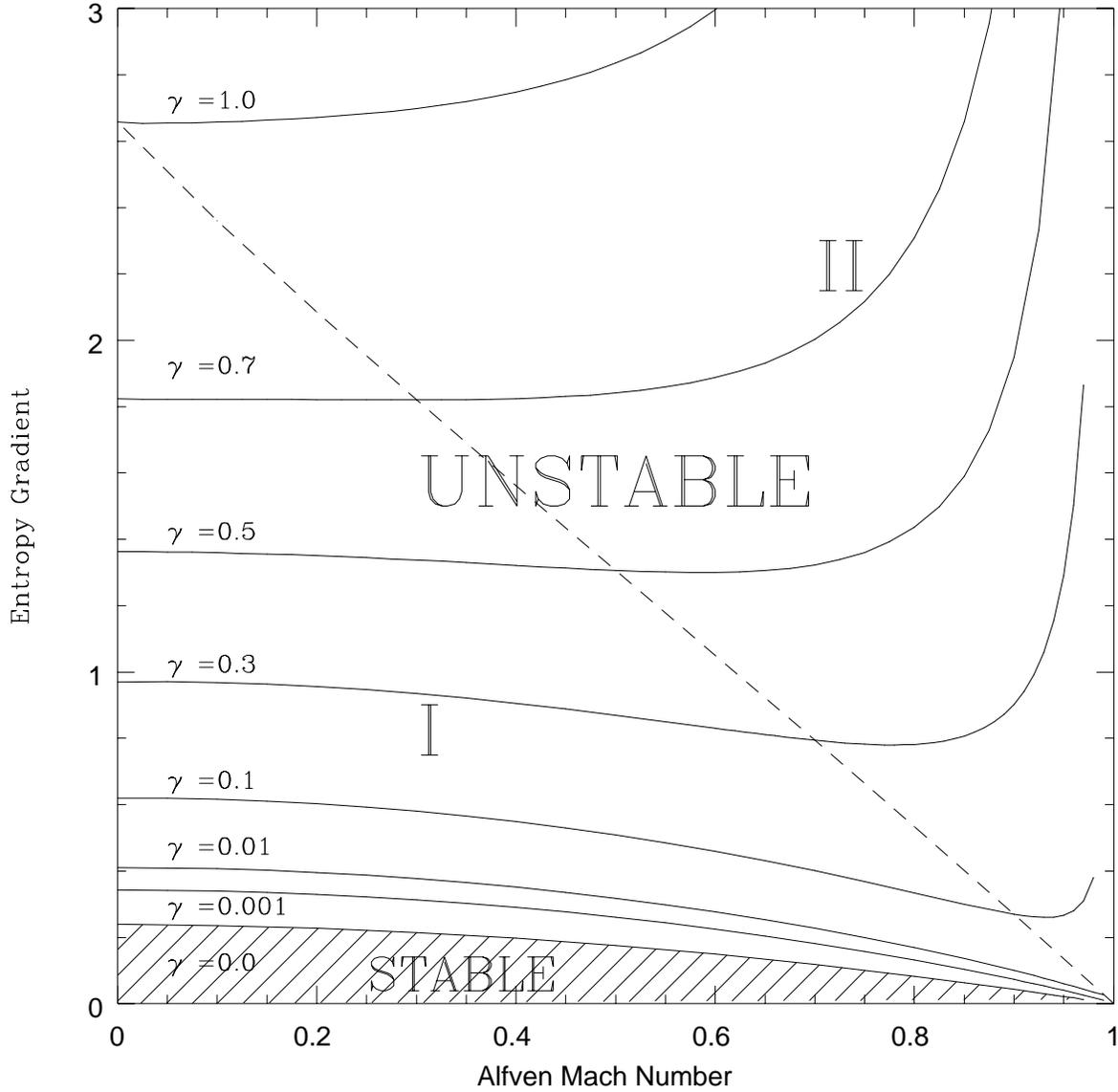}
\caption{Stability diagram for the Boussinesq limit:
Contours of constant normalized growth rate $\gamma$ are plotted over
the space of entropy gradient $s_0'$ vs. Alfv\'{e}n Mach number $M$.
The stable parameter regime is denoted by hashing. The diagonal dotted
line denotes $\gamma=1-M$, separating unstable region (I), where flow
enhances instability growth, from unstable region (II), where flow
suppresses the instability.}
\label{fig1}
\end{figure}

\begin{figure}
\plotone{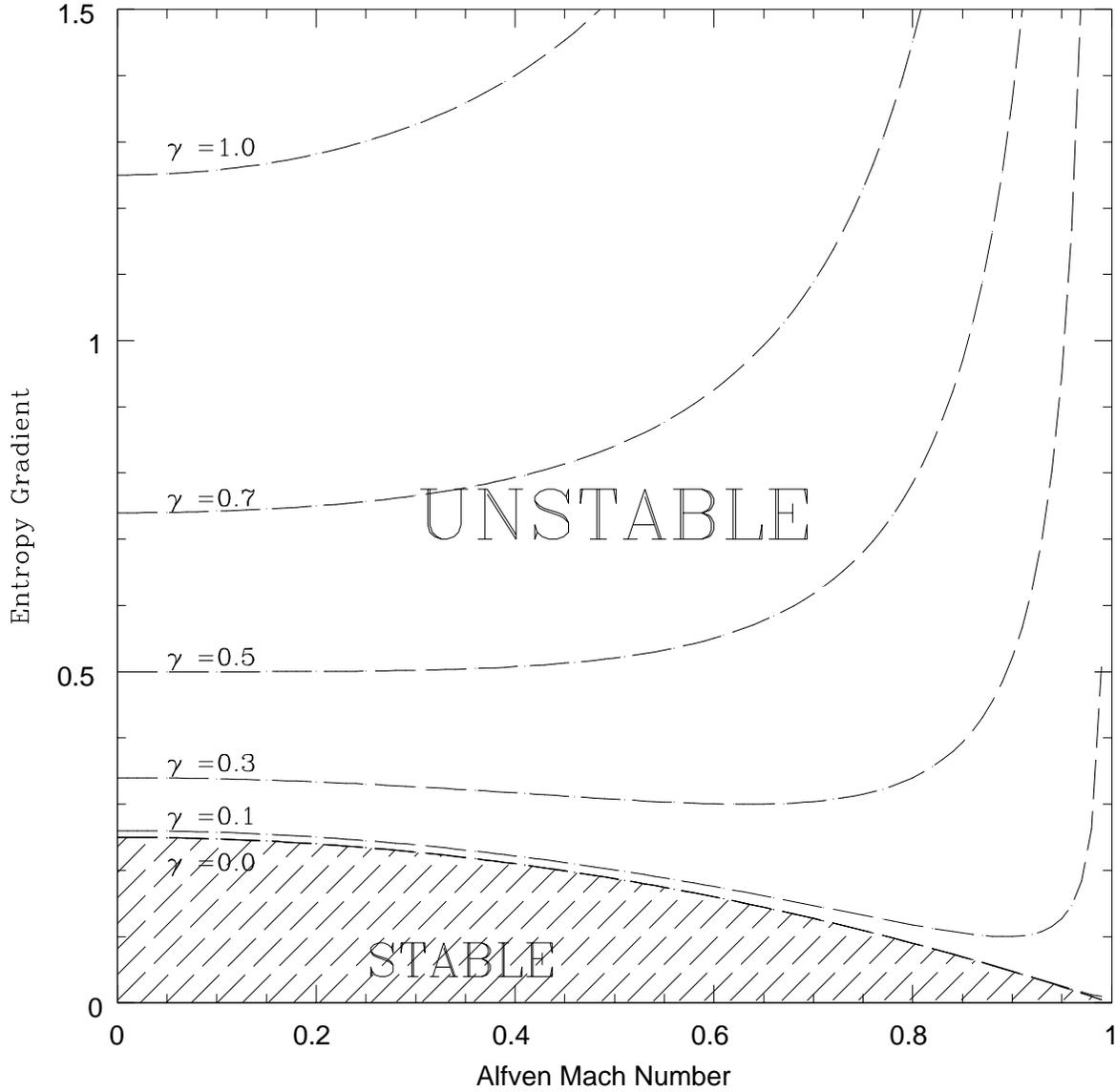}
\caption{The stability diagram for the straight-field case as 
described by equation~(\ref{eq:simpeig}) in section~\ref{sec:str8}.
Note that the qualitative features of the diagram are similar to those
in figure~\ref{fig1}.  A value of $L=\pi$ was chosen to plot this
diagram.}
\label{fig:simp}
\end{figure}

\begin{figure}
\plottwo{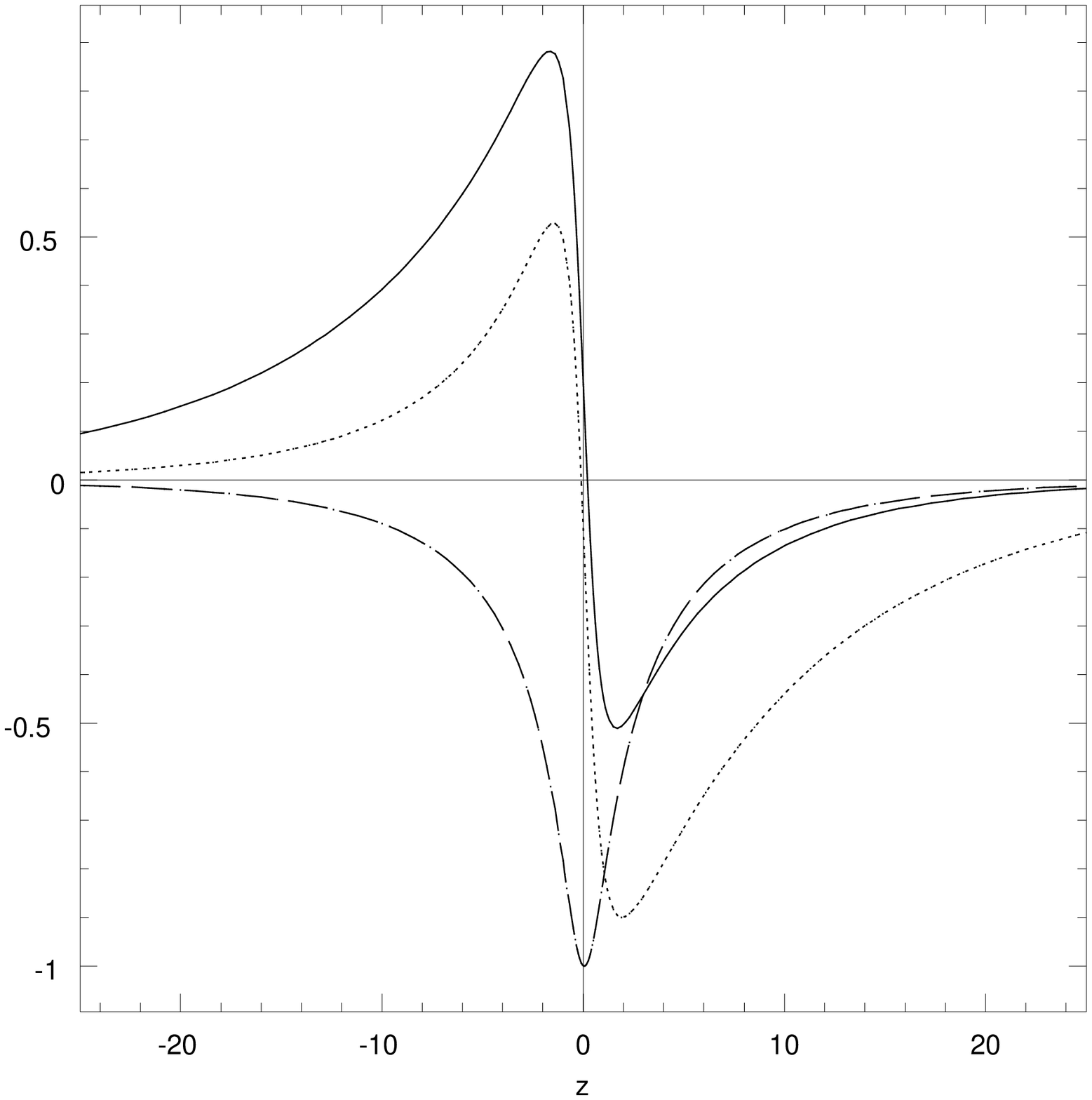}{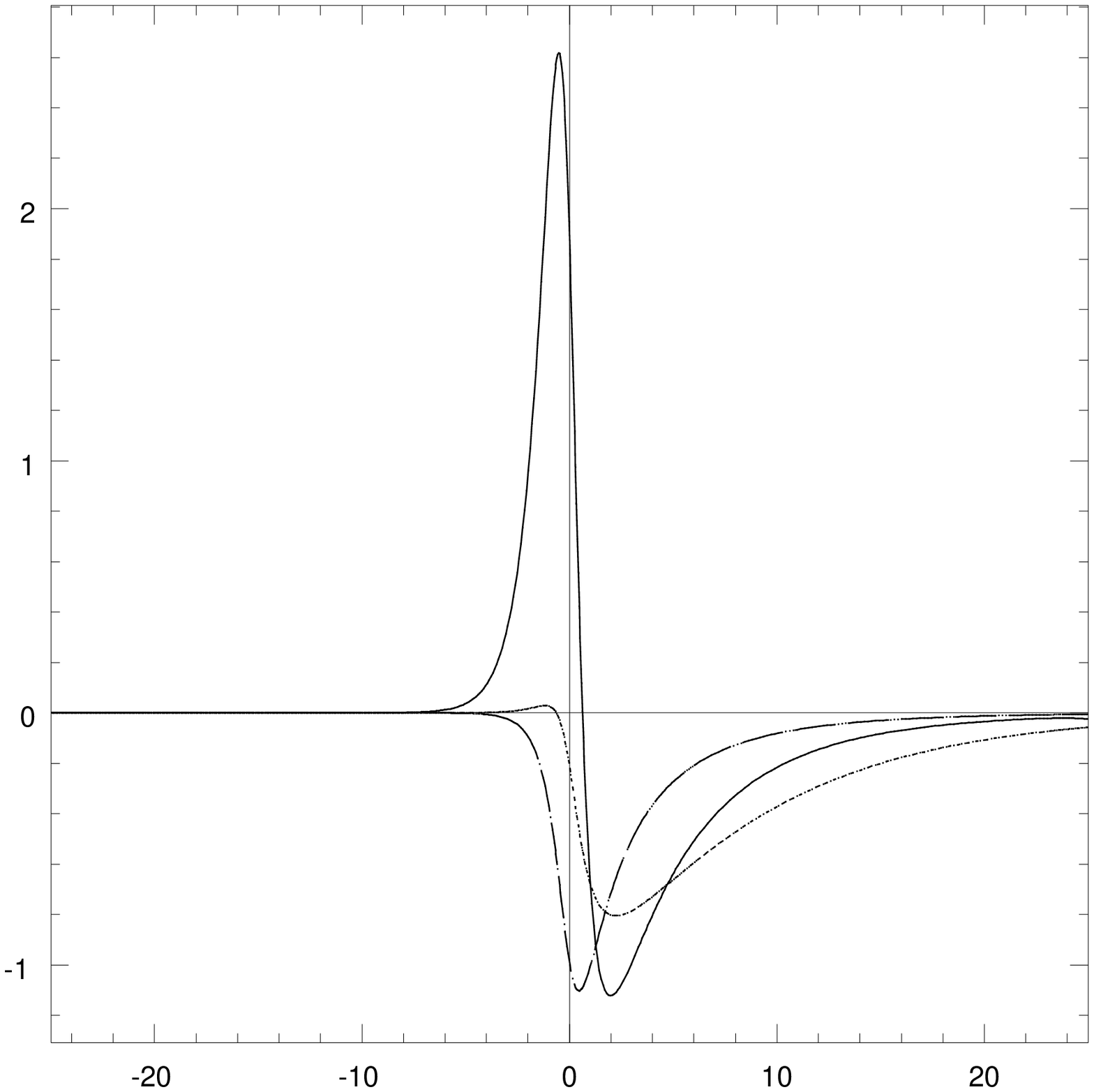}
\caption{Eigenfunctions for entropy gradient $s_0'=0.6$ 
and $M=0$ (left plot) and $M=0.8$ (right plot).  The three functions 
are $A_+$ (solid line), $A_-$ (dotted line), and $s$ (dashed line).  Note 
that for the $M=0.8$ case, where the plasma flow is in the $+z$ direction, 
the eigenfunctions grow quickly with steep gradients and diminish slowly
as you move from left to right.}
\label{fig:eigfunc}
\end{figure}

\begin{figure}
\plotone{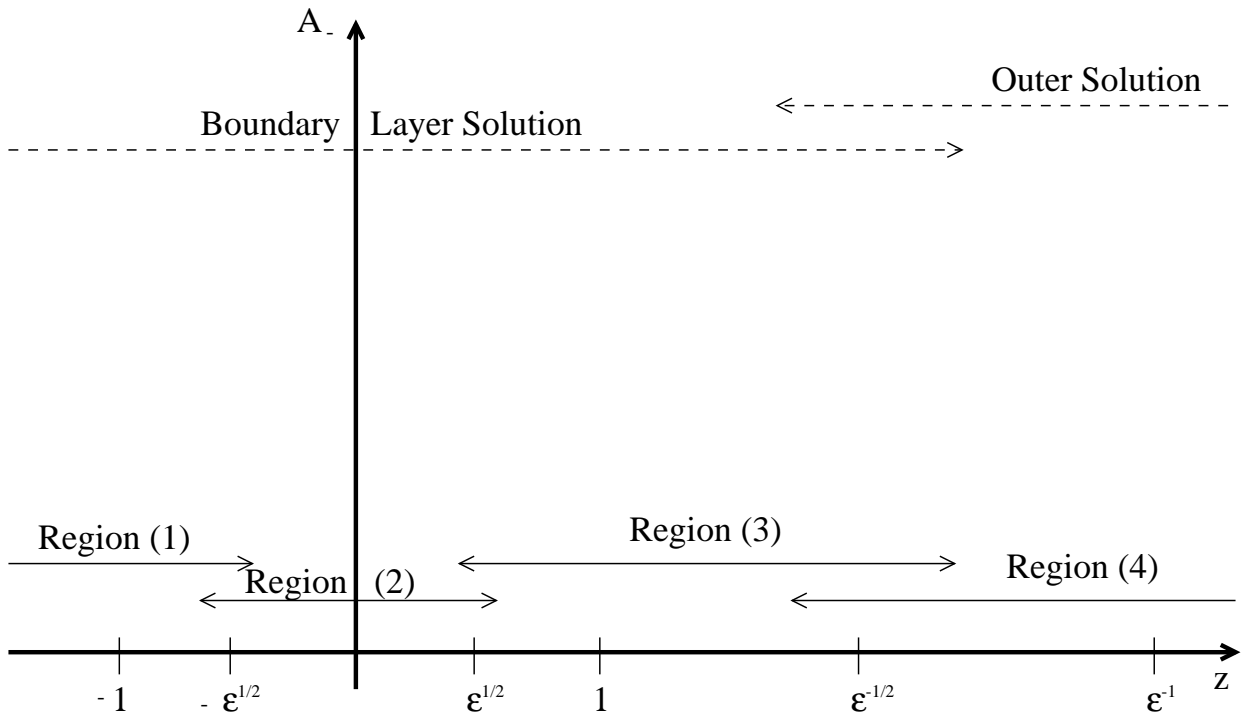}
\caption{Regions defined for the asymptotic solution of 
$A_-$ in the limit $M \rightarrow 1$.}
\label{fig:asymp}
\end{figure}

To demonstrate stabilization as the Alfv\'en Mach
number approaches one, as seen of region (II) of Figure~\ref{fig1}, we
perform an asymptotic analysis of our model in the limit $M
\rightarrow 1$.  For a small dimensionless parameter $\epsilon$, we
quantify the order of this limit as $1-M \sim
\mathcal{O}(\epsilon^2)$. We expect the instability growth rate to be
$\gamma \sim \mathcal{O}(\epsilon)$.  Identifying the terms in the
dimensionless system of equations~(\ref{eq:A+b})--(\ref{eq:sb}) for
reference, we have
\begin{equation}
\begin{array}{ccccc}
 - ( 1 - M ) \frac{d A_{+}}{d z} & =  & - \gamma A_{+} &+  ( 1 + M ) 
	\frac{z}{1 + z^2} A_{-} &- \frac {s}{(1 + z^2)^{1/2}} \label{eq:A+} \\
 (1) & & (2) & (3) & (4) \nonumber 
\end{array} 
\end{equation}
\begin{equation}
\begin{array}{ccccc}
 ( 1 + M ) \frac{d A_{-}}{d z}  & =  &- \gamma A_{-} &-   ( 1 - M ) 
	\frac{z}{1 + z^2} A_{+} &+ \frac {s}{(1 + z^2)^{1/2}} \label{eq:A-} \\
 (5) & & (6) & (7) & (8) \nonumber 
\end{array} 
\end{equation}
\begin{equation}
\begin{array}{ccccc}
 M \frac{d s}{d z} &=&  - \gamma s & - \frac{s_0'}{2} \frac{A_{+}}
	{(1 + z^2)^{1/2}} & + \frac{s_0'}{2}\frac{A_{-} }{(1 + z^2)^{1/2}} .
\label{eq:s} \\
 (9) & & (10) & (11) & (12) \nonumber 
\end{array} 
\end{equation}
The boundary conditions demand $A_+, A_-, s \rightarrow 0$ as $|z|
\rightarrow \infty$.  Asymptotic solutions in the $M \rightarrow 1$
limit can be found in the four regions along $z$ displayed in
Figure~\ref{fig:asymp}.  Below we find the solutions for each of these
regions and, by matching the solutions between these regions, we
obtain an eigenvalue condition on the growth rate demonstrating
stabilization as the Alfv\'en Mach number approaches one.  First, we
obtain two reductions of equations~(\ref{eq:A+})--(\ref{eq:s}); one
over a boundary layer where $|z| \ll \epsilon^{-1}$, and the other
over an outer region where $|z| \gg 1$ .  Then, we present the
solutions in each of the four regions in Figure~\ref{fig:asymp}.

As the Alfv\'en Mach number approaches one, regions (1), (2), and (3)
of Figure~\ref{fig:asymp} behave like a boundary layer region: we
expect derivatives to be large and thus take $\frac{d}{dz} \sim
\mathcal{O}(\epsilon^{-1})$.  We treat $z\sim \mathcal{O}(1)$ over
these three regions.  Balance of the dominant terms (9) and (11) in
equation~(\ref{eq:s}) shows that $s \sim \mathcal{O}(\epsilon A_+)$.
In turn, this demands that terms (5) and (8) must balance in
equation~(\ref{eq:A-}), yielding the ordering $A_- \sim
\mathcal{O}(\epsilon^2 A_+)$.  Adopting the specified ordering allows
us to drop terms (6), (7), (10), and (12) from
equations~(\ref{eq:A+})--(\ref{eq:s}); term (3), although one order in
$\epsilon$ smaller than the other terms in equation~(\ref{eq:A+}),
will contribute in the regions for $z > 1$, so we retain it in order
to be certain that our first order correction in region (3) is valid.
The remaining equations can be combined to a single third-order
equation in $A_-$ and simplified by the substitution $z= \sinh \theta$
to obtain
\begin{equation}
\frac{2 M(1-M)}{s_0'}\frac{d^3 A_-}{d \theta^3} - \frac{2 \gamma M}{s_0'}
\cosh \theta \frac{d^2 A_-}{d \theta^2} 
+ \frac{d A_-}{d \theta} - \tanh \theta A_- = 0.
\label{eq:a1}
\end{equation}

In region (4), we find the smoothly varying outer solution over which
$z \sim \mathcal{O}(\epsilon^{-1})$. We expect that $\frac{d}{dz}
\sim \mathcal{O}(\epsilon)$ here.  By insisting that terms (9) and (10)
balance with term (11) in equation~(\ref{eq:s}), we find that $s \sim
\mathcal{O}(A_+)$.  Similarly, by balancing terms (5) and (6) with
term (8) in equation~(\ref{eq:A-}), we obtain $A_- \sim
\mathcal{O}(A_+)$.  This ordering allows us to drop terms (1) and (7)
in equations~(\ref{eq:A+})--(\ref{eq:s}).  Approximating
$(1+z^2)^{1/2} \simeq z$, substituting $A_-= z B_-$, and combining the
equations, we obtain the second-order equation
\begin{eqnarray}
M(1+M) z^2 \frac{d^2 B_-}{dz^2} &+ \left[ (1 + 2 M) \gamma z^2 + 3 M(1+M) z 
- \frac{(1+M) s_0'}{2 \gamma} \right] \frac{d B_-}{dz}  \nonumber \\
&+ \left[ \gamma^2 z^2 + (1 + 3 M)\gamma z + M(1+M) - s_0' \right] B_- =0.
\label{eq:a2}
\end{eqnarray}

To find the solution in region (1), we assume an eikonal solution for
equation~(\ref{eq:a1}).  Neglecting the trivial constant solution, we find
two independent solutions of the form
\begin{equation}
 A_- \sim \frac{1}{(1+z^2)^{3/4}\left[(1+z^2)^{1/2} +z\right]^{1/2}} 
\exp \left( \frac{\gamma }{2(1-M)}  \left[z + (1+z^2)^{1/2}\right]\right)
 \label{eq:a-1a}
\end{equation}
\begin{equation}
 A_- \sim \frac{z + (1+z^2)^{1/2}}{(1+z^2)^{1/2}} 
\exp \left(\frac{\gamma }{2(1-M)} \left[z - (1+z^2)^{1/2}\right] + z\left[z-(1+z^2)^{1/2}\right] \right).
\label{eq:a-2a}
\end{equation}
To get the behavior for $|z| \gg 1$, we can expand $(1+z^2)^{1/2}
\simeq |z| + 1/2|z|$.  For region (1), we note that $z<0$ and,
retaining only the dominant terms, we obtain solutions of the form
\begin{equation}
 A_- \sim \frac{1}{z} \exp \left( \frac{-\gamma }{4(1-M)z} \right)
 \label{eq:a-1reg1}
\end{equation}
\begin{equation}
 A_- \sim\frac{1}{z^2}  \exp \left( \frac{\gamma z}{1-M} \right).
\label{eq:a-2reg1}
\end{equation}
The boundary conditions impose that $A_- \rightarrow 0$ as $z
\rightarrow -\infty$, so our solution in region (1) must be entirely
of the form of equation~(\ref{eq:a-2reg1}), a growing solution in the
$+z$ direction. To determine the behavior of 
equations~(\ref{eq:a-1a}) and (\ref{eq:a-2a}) in the overlap with
region (2), we take the limit $|z| \ll 1$ and approximate
$(1+z^2)^{1/2} \simeq 1 + z^2/2$ to get the two solutions
\begin{equation}
 A_- \sim \exp \left( \frac{\gamma }{2(1-M)} \left[z+z^2/2\right] \right)
 \label{eq:a-1reg1small}
\end{equation}
\begin{equation}
 A_- \sim \exp \left( \frac{\gamma}{2(1-M) }\left[z-z^2/2\right] \right).
\label{eq:a-2reg1small}
\end{equation}
One of these solutions must smoothly match onto the solution for region (2).

Region (1)'s solution will be valid as we move in the
$+z$ direction until the eikonal approximation,
$\frac{\gamma z}{1-M} \gg 1$, breaks down.  The failure of this
condition occurs in region (2) of Figure~\ref{fig:asymp} where $z
\sim
\mathcal{O}(\epsilon)$. For region (2), we expand
equation~(\ref{eq:a1}) about $z=0$.  For $z \ll 1$, $\cosh \theta
\simeq 1$ and $\tanh \theta \simeq \theta$ where $\theta \ll 1$, so we
can drop the last term of equation~(\ref{eq:a1}).  Letting $f =
\frac{dA_-}{d \theta}$ and using the integrating factor $f= B_- \exp
\left( \frac{ \gamma \sinh \theta }{2 (1-M)} \right)$ to simplify the
result, we obtain the equation
\begin{equation}
\frac{d^2 B_-}{d \theta^2} - \left[ \frac{\gamma^2  \cosh^2 \theta}{4(1-M)^2}- 
\frac{  \gamma \sinh \theta}{2(1-M)}
 -\frac{s_0'}{2 M (1-M)}  \right] B_- =0.
\label{eq:B-HO}
\end{equation}
Neglecting the central term in the coefficient of $B_-$ because it is
an order $\epsilon$ smaller than the other terms, we can cast
equation~(\ref{eq:B-HO}) in the form of Hermite's equation for which
the solutions are well known.  Therefore, for the $n=0$ Hermite
polynomial, the solution in region (2) is
\begin{equation}
 A_- \sim \int^z \exp \left( \frac{\gamma }{2(1-M)} (z' - z'^2/2) \right) dz',
\label{eq:a-reg2}
\end{equation}
and the eigenvalue condition on the growth rate imposed by Hermite's
equation is
\begin{equation}
\gamma^2 = \frac{2 s_0' (1-M)}{M} - 2 (2n+1) \gamma (1-M)
\label{eq:eigcond}
\end{equation}
for the $n$th Hermite polynomial.  Thus, we find that the solution in
region (1) given by equation~(\ref{eq:a-2reg1small}) matches smoothly
onto our solution in region (2).

In region (3), we once again assume an eikonal solution for
equation~(\ref{eq:a1}) and find the two solutions given by
equations~(\ref{eq:a-1a}) and (\ref{eq:a-2a}).  To match with region
(2), find the $|z| \ll 1$ limit of these equations, yielding once more
equations~(\ref{eq:a-1reg1small}) and (\ref{eq:a-2reg1small}); we
observe that equation~(\ref{eq:a-2reg1small}) for the small $z$ limit
of region (3) matches solution~(\ref{eq:a-reg2}) for region (2). In the $|z| \gg1$
limit of equations~(\ref{eq:a-1a}) and (\ref{eq:a-2a}) for region (3),
we obtain the solutions (for $z>0$)
\begin{equation}
 A_- \sim\frac{1}{z^2}  \exp \left( \frac{\gamma z}{1-M} \right)
 \label{eq:a-1reg3}
\end{equation}
\begin{equation}
 A_- \sim \exp \left( \frac{-\gamma }{4(1-M)z} \right).
\label{eq:a-2reg3}
\end{equation}
To continue our
asymptotic solution, we must smoothly match one of these region (3)
solutions to the solution for region (4) in the overlap around $z
\sim \mathcal{O}(\epsilon^{-1/2})$.

In region (4), we assume eikonal solutions for equation~(\ref{eq:a2})
in the limit $z \rightarrow 0$.  The two solutions found are
\begin{equation}
 A_- \sim z
 \label{eq:a-1reg4}
\end{equation}
\begin{equation}
 A_- \sim \exp \left( \frac{-s_0' }{2 \gamma M z} \right).
\label{eq:a-2reg4}
\end{equation}
Hence, we can match the solution given by equation~(\ref{eq:a-2reg3})
in region (3) with the solution given by equation~(\ref{eq:a-2reg4})
in region (4) if
\begin{equation}
\frac{\gamma}{4 (1-M)}  = \frac{s_0'}{2 \gamma M}.
\label{eq:matchcond}
\end{equation}
But, this is identical to the lowest order of the eigenvalue
condition, equation~(\ref{eq:eigcond}). To complete our asymptotic
solution, we must find that a solution to equation~(\ref{eq:a2}) in
the limit $z \rightarrow \infty$ which satisfies the boundary
condition that $A_- \rightarrow 0$ as $z \rightarrow \infty$.  In this
limit, the two solutions take the form
\begin{equation}
 A_- \sim \exp \left( \frac{-\gamma z }{M} \right)
 \label{eq:a-1reg4inf}
\end{equation}
\begin{equation}
 A_- \sim \exp \left( \frac{-\gamma z }{1+M} \right).
\label{eq:a-2reg4inf}
\end{equation}
Both of these solutions satisfy the boundary condition as $z
\rightarrow \infty$.

Now that we have seen that it is possible to construct a complete
asymptotic solution in the limit $M \rightarrow 1$, let us examine
this solution more closely.  Beginning in region (1) at the left of
Figure~\ref{fig:asymp}, the boundary conditions demand that the
solution must be solely of the form of
equation~(\ref{eq:a-2reg1}). But, as behavior in regions (1), (2), and
(3) is governed by equation~(\ref{eq:a1}), the eikonal approximation
must break down in region (2) in order for the solution in region (1)
to convert to the solution given by equation~(\ref{eq:a-2reg3}) in
region (3) so that smooth matching may be accomplished with
solution~(\ref{eq:a-2reg4}) in region (4).  The failure of the eikonal
approximation around $z=0$ yields a reduction of
equation~(\ref{eq:a1}) to the Hermite-type equation~(\ref{eq:B-HO}).
The requirement that a solution to this equation exist imposes the
eigenvalue condition, equation~(\ref{eq:eigcond}).  This single
condition can also be used to smoothly match
solution~(\ref{eq:a-2reg3}) in region (3) to
solution~(\ref{eq:a-2reg4}) in region (4).  Finally, region (4) is
governed by equation~(\ref{eq:a2}).  In the limit $z \rightarrow 0$,
this equation yields a matching solution in the overlap with region
(3); and, in the limit $z \rightarrow \infty$, it provides two
solutions which both satisfy the boundary conditions as $z \rightarrow
\infty$.  Therefore, the single condition necessary to find a smooth
solution which satisfies the boundary conditions is the eigenvalue
condition, equation~(\ref{eq:eigcond}).  To lowest order, this
condition can be written in a more recognizable form in the limit $M
\rightarrow 1$ as
\begin{equation}
\gamma^2 \simeq s_0' (1-M^2).
\label{eq:eigcond2}
\end{equation}
Thus,  in region (II) of Figure~\ref{fig1}, where $\gamma \gg 1-M$, the behavior is clearly demonstrated---that stabilization occurs as the Alfv\'en Mach
number approaches one.


\clearpage

\begin{thebibliography}{}
\bibitem[Adam(1978)]{ada78} Adam, J. A. 1978, J. Plasma Phys., 19, 77.
\bibitem[Balbus and Hawley(1991)]{bal91} Balbus, S. A. and Hawley, J. F. 
	 1991, \apj, 376, 214.
\bibitem[Barnes et. al.(1998)]{bar98} Barnes, G., MacGregor, K. B., 
	and Charbonneau, P. 1998, \apj, 498, L169.
\bibitem[Bender and Orszag(1978)]{ben78} Bender, C. M. and Orszag, S. A. 
	1978, Advanced Mathematical Methods for Scientists and Engineers,
	New York: McGraw-Hill.
\bibitem[Bernstein et. al.(1958)]{ber58} Bernstein, I. B., Frieman, E. A., 
	Kruskal, M. D., and Kulsrud, R. M. 1958, Proc. Roy. Soc., A244, 17.
\bibitem[Blaes and Balbus(1994)]{bla94} Blaes, O. M. and Balbus, S. A.  
	 1994, \apj, 421, 163.
\bibitem[Caligari et. al.(1995)]{cal95} Caligari, P., Moreno-Insertis, F. 
	and Schussler, M. 1995,  \apj, 441, 886.
\bibitem[Cally(1983)]{cal83a} Cally, P. S. 1983, Geophys. 
	Astrophys. Fluid Dynamics, 23, 43.
\bibitem[Cally and Adam(1983)]{cal83b} Cally, P. S. and Adam, J. A. 1983,
	Geophys. Astrophys. Fluid Dynamics, 23, 57.
\bibitem[Cattaneo and Hughes(1988)]{cat88} Cattaneo, F.
	and Hughes, D.W.  1988, J. Fluid Mech., 196, 323.
\bibitem[Cattaneo et. al.(1990)]{cat90} Cattaneo, F., Chiueh, T., 
	and Hughes, D.W.  1990, J. Fluid Mech., 219, 1.
\bibitem[Centrella and Wilson(1984)]{cen84} Centrella, J. and Wilson, J. R. 
	  1984, \apjs, 54, 229.
\bibitem[Chandrasekhar(1960)]{cha60} Chandrasekhar, S. 1960, Proc. Natl. Acad.
	Sci., 46, 253.
\bibitem[Chandrasekhar(1961)]{cha61} Chandrasekhar, S. 1961, Hydrodynamic 
	and Hydromagnetic Stability , Oxford: Oxford Univ. Press.
\bibitem[Charbonneau and MacGregor(1996)]{cha96} Charbonneau, P. and 
	MacGregor, K. B. 1997, \apj, 473, L59.
\bibitem[Cowley et. al.(1991)]{cow91} Cowley, S. C., Kulsrud, R.M. and  
	Sudan, R.  1991, Phys. Fluids B, 3, 2767.
\bibitem[Gilman and Fox(1997)]{gil97} Gilman, P. A. and Fox P. A. 1997, 
	\apj, 484, 439.
\bibitem[Goldreich and Lynden-Bell(1965)]{gol65} Goldreich, P. and 
	Lynden-Bell, D. 1965, \mnras, 130, 125.
\bibitem[Hassam(1991)]{has91} Hassam,  A. B. 1991, Comments Plasma Phys. 
	and Controlled Fusion, 14, 275.
\bibitem[Hassam(1996)]{has96} Hassam,  A. B. 1996, Nuclear Fusion, 36, 707.
\bibitem[Hassam(1999)]{has99} Hassam,  A. B. 1999, Phys. Plasmas, 6, 3772.
\bibitem[Hayashi and Young(1987)]{hay87} Hayashi, Y.-Y. and Young, W. R. 
	1987, J. Fluid Mech., 184, 477.
\bibitem[Hughes and Cattaneo(1987)]{hug87} Hughes, D.W. 
	and Cattaneo, F. 1987, Geophys. Astrophys. Fluid Dynamics, 39, 65.
\bibitem[Ince(1926)]{inc26} Ince, E. L.  1926, Ordinary Differential 
    	Equations, New York: Dover.
\bibitem[Kitchatinov and R\"udiger(1997)]{kit97} Kitchatinov, L.
	 and R\"udiger, G. 1997, \mnras, 286, 757.
\bibitem[Miller et. al.(1995)]{mil95} Miller, R. L., Waelbroeck, F. L., 
	Hassam, A. B., and Waltz, R. E. 1995, Phys. Plasmas, 2, 3676.
\bibitem[Newcomb(1961)]{new61} Newcomb, W. A. 1961, Phys. Fluids, 4, 391.
\bibitem[Parker(1979)]{par79} Parker, E. N.  1979, Cosmical Magnetic Fields: 
	Their Origin and thier Activity, Oxford: Clarendon Press.
\bibitem[Parker(1993)]{par93} Parker, E. N. 1993, \apj, 408, 707.
\bibitem[Roberts and Taylor(1965)]{rob65} Roberts, K. V. and Taylor, J. B.
	 1965, Phys. Fluids, 8, 315.
\bibitem[Schou et. al.(1998)]{sch98} Schou, J. et. al. 1998, \apj, 505, 390.
\bibitem[Schussler et. al.(1994)]{sch94} Schussler, M., Caligari, P., 
	Ferriz-Mas, A. and Moreno-Insertis, F.  1994,  Astron. Astrophys., 
	281, L69.
\bibitem[Schwarzschild(1906)]{sch06} Schwarzschild, K.   1906, Nachr. 
	Kgl. Ges. Wiss. G\"{o}ttingen, p. 41.
\bibitem[Spiegel and Zahn(1992)]{spi92} Spiegel, E. A. and Zahn, J.-P. 
	1992, Astron. Astrophys., 265, 106.
\bibitem[Thompson(1996)]{tho96} Thompson, M. J. et. al. 1996, Science, 272, 
	1300.
\bibitem[Urpin(1996)]{urp96} Urpin, V. 
	1996, \mnras, 280, 149.
\bibitem[Velikhov(1959)]{vel59} Velikhov, E. P. 1959, Sov. Phys. JETP,
	36, 995.
\bibitem[Waelbroeck and Chen(1991)]{wae91} Waelbroeck, F. L., and Chen, L. 
	1991, Phys. Fluids B, 3, 601.
\bibitem[Zweibel and Heiles(1997)]{zwe97} Zweibel, E. G., and Heiles, C.
	 1997, Nature, 385, 131.


\end{thebibliography}
\end{document}